\documentclass[a4paper]{jpconf}
\usepackage{graphicx}
\begin{document}

\newcommand{\mnv}{\mathrm{MINER}{\nu}\mathrm{A}}

\title{$\mnv$ Status and Event Reconstruction}

\author{Gabriel N. Perdue, on behalf of the $\mnv$ Collaboration}

\address{University of Rochester, Rochester, New York 14610 USA}

\ead{perdue@fnal.gov}

\begin{abstract}
$\mnv$ (Main INjector ExpeRiment $\nu$-A) is a few-GeV neutrino scattering experiment that began taking data in the NuMI beam at Fermilab (FNAL) in the Fall of 2009.  $\mnv$ employs a fine-grained detector, with an eight ton active target region composed of plastic scintillator. It also uses nuclear targets composed of carbon, iron, and lead placed upstream of the active region to measure $\nu$-A dependence.  The experiment will provide important inputs for neutrino oscillation experiments and a pure weak probe of nuclear structure.  We offer a set of initial kinematic distributions of interest and provide a summary of current operations and event reconstruction status.
Contribution to NUFACT 11, XIIIth International Workshop on Neutrino Factories, Super beams and Beta beams, 1-6 August 2011, CERN and University of Geneva
(Submitted to IOP conference series).
\end{abstract}

\section{Introduction}\label{sec:intro}

$\mnv$ is an on-axis neutrino-nucleus scattering experiment at Fermilab's NuMI (Neutrinos at the Main Injector) beamline that will measure interaction cross-sections and event kinematics in exclusive and inclusive states to high precision. 
The experiment will also examine nuclear effects and parton distribution functions (PDF's) using a variety of targets materials.

\subsection{Precision Neutrino Oscillation Experiments and Nuclear Effects}\label{subsec:precneut}
Interaction details are crucial for neutrino energy estimation and in background separation in  oscillation experiments. 
For example, in a $\nu_{\mu}$ disappearance experiment, the two-flavor disappearance relation is show in Eq. \ref{eq:numudisp}:
\begin{equation}
P\left( \nu_{\mu} \rightarrow \nu_{\mu} \right) = 1 - \sin^2 \left( 2\theta_{23} \right) \sin^2 \left( \frac{1.27 \Delta m_{23}^2 (eV^2) L(km)}{E_{\nu}(GeV)} \right)
\label{eq:numudisp}
\end{equation}
Experiments cannot directly measure the neutrino energy that appears in this relation, and visible energy is a function of flux, cross-section, and detector response. 
Interactions occur in dense nuclear matter, making final state interactions (FSI) significant in producing the observed particles.
Near-to-Far Detector ratios cannot entirely eliminate the associated uncertainties because the energy spectra at the two detectors are different due to beam, oscillation, matter, and even nuclear effects if the detector materials differ.

\begin{figure}[h]
\includegraphics[width=16pc]{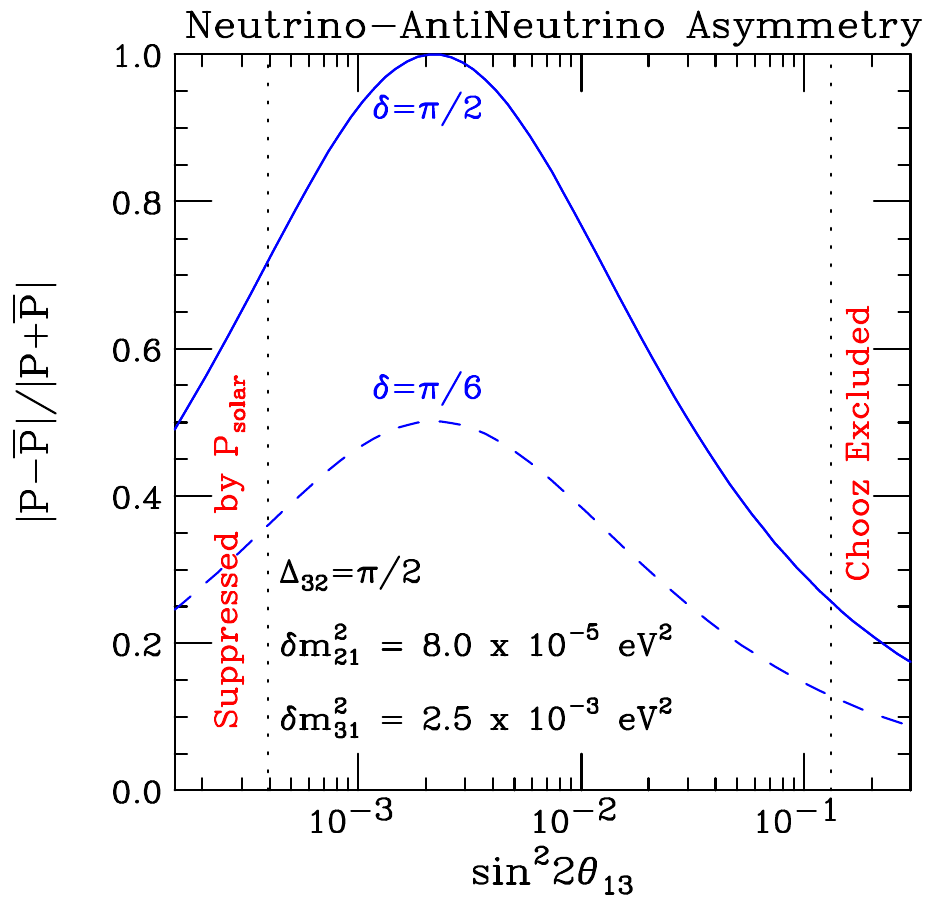}\hspace{2pc}
\begin{minipage}[b]{14pc}\caption{\label{fig:nunubarAsym}Oscillation asymmetry between neutrinos and anti-neutrinos as a function of $\sin^2 2\theta_{13}$ for different CP violating phases. Figure adapted from \cite{cpvio}.}
\end{minipage}
\end{figure}

Understanding these effects is especially important in light of recent ``hints'' of a large $\theta_{13}$ \cite{t2k}. As Figure \ref{fig:nunubarAsym} illustrates, for a fixed value of the CP violating phase, $\delta$, the $\nu$-$\bar{\nu}$ oscillation asymmetries are smaller at larger values of $\theta_{13}$. Therefore, for larger values of $\theta_{13}$, experiments are measuring a smaller difference between two larger numbers and systematic errors increase in relative importance. 
In this case, $\mnv$ cross-section and kinematic measurements increase in value.

Additionally, $\mnv$ supports a nuclear effects program complementary to charged lepton scattering measurements. Many quantities of interest carry large uncertainties: axial form factors as a function of A and momentum transfer ($Q^2$), quark-hadron duality, $x$-dependent nuclear effects, etc.

\section{The $\mnv$ Detector and Operations}\label{sec:detlive}

\begin{figure}
\begin{center}
\includegraphics[width=28pc]{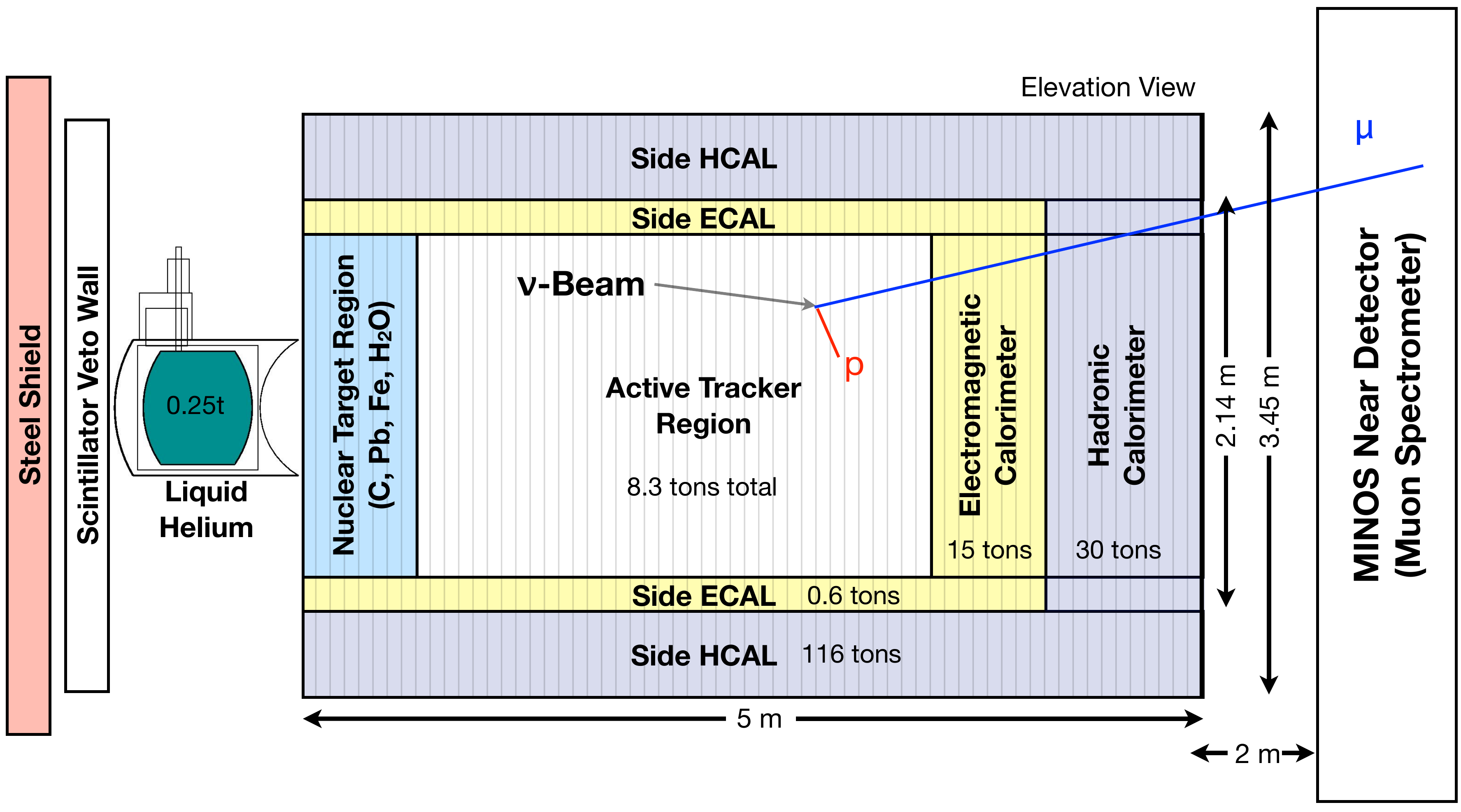}
\end{center}
\caption{\label{fig:detector}The $\mnv$ detector as of Summer 2011. (The water target is not installed yet.)}
\end{figure}

$\mnv$ is a horizontal stack of 120 similar modules weighing approximately two tons each. 
Each module has an inner detector (ID) composed of triangular plastic scintillator strips and an outer detector (OD) steel frame instrumented with plastic scintillator bars. 
Most modules feature an ID composed of two stereoscopic planes of scintillator, but some in the nuclear target and calorimetric regions of the detector give up one or both scintillator planes for target or absorber materials. 
See Figure \ref{fig:detector} for a schematic of the detector.

\begin{figure}[ht]
\begin{minipage}[b]{0.6\linewidth}
\includegraphics[width=20pc]{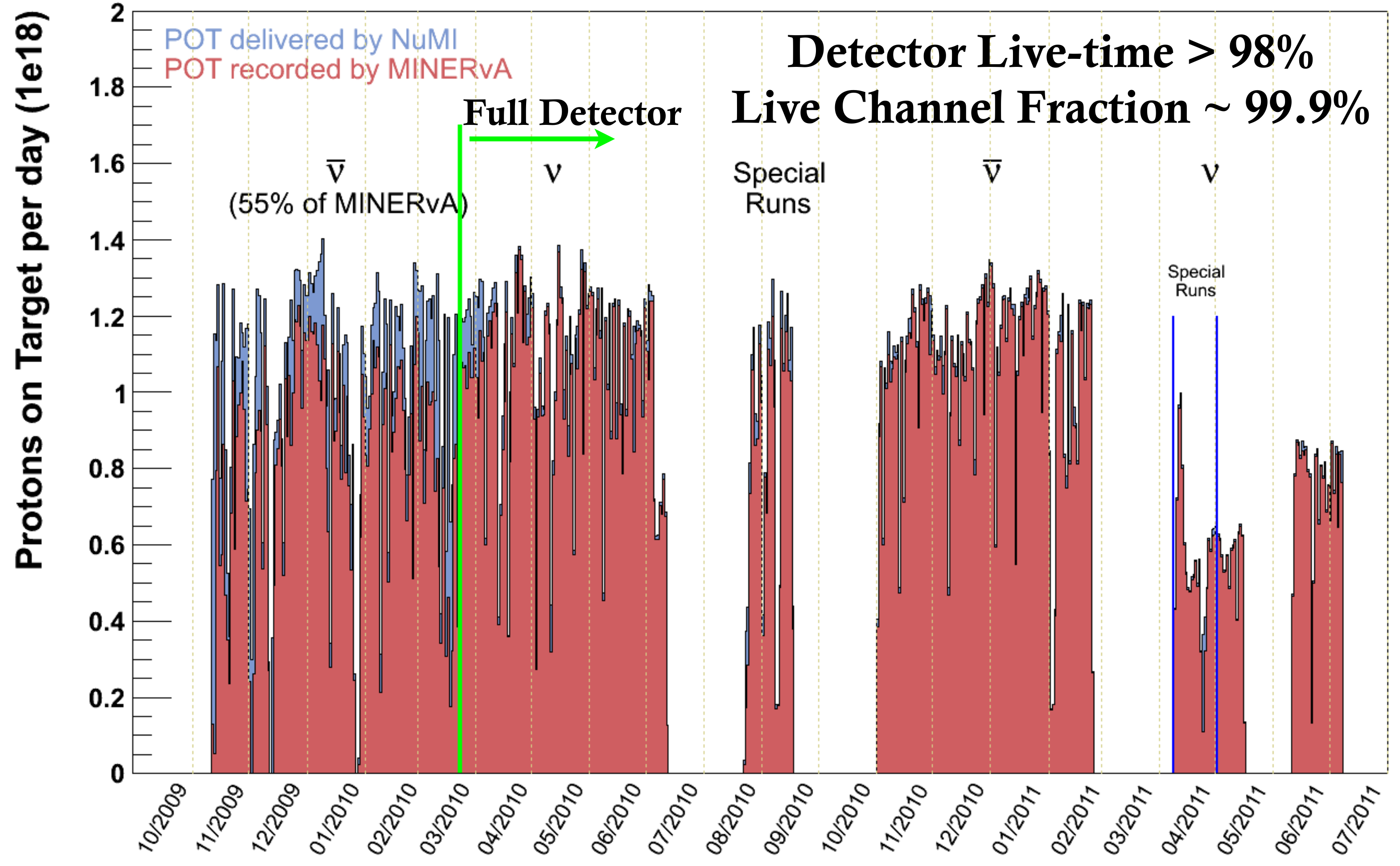}\hspace{2pc}
\end{minipage}
\begin{minipage}[b]{0.4\linewidth}
\caption{\label{fig:potlive}Accumulated protons on target (P.O.T.) and live-time. Prior to March 2010, we integrated data using a partially constructed version of the detector.}
\end{minipage}
\end{figure}

Figure \ref{fig:potlive} illustrates integrated data collection and livetime.
$\mnv$ began its physics run in March, 2010 in the NuMI beam forward horn current (FHC) mode, focusing $\pi^{+}$ mesons (``neutrino mode'').  Reverse horn current (RHC) mode data, focusing $\pi^{-}$ mesons (``anti-neutrinos mode''), was taken prior to March, 2010 and again in Winter 2010-2011. 

\section{Particle ID}\label{sec:pid}

\begin{figure}
\begin{center}
\includegraphics[width=28pc]{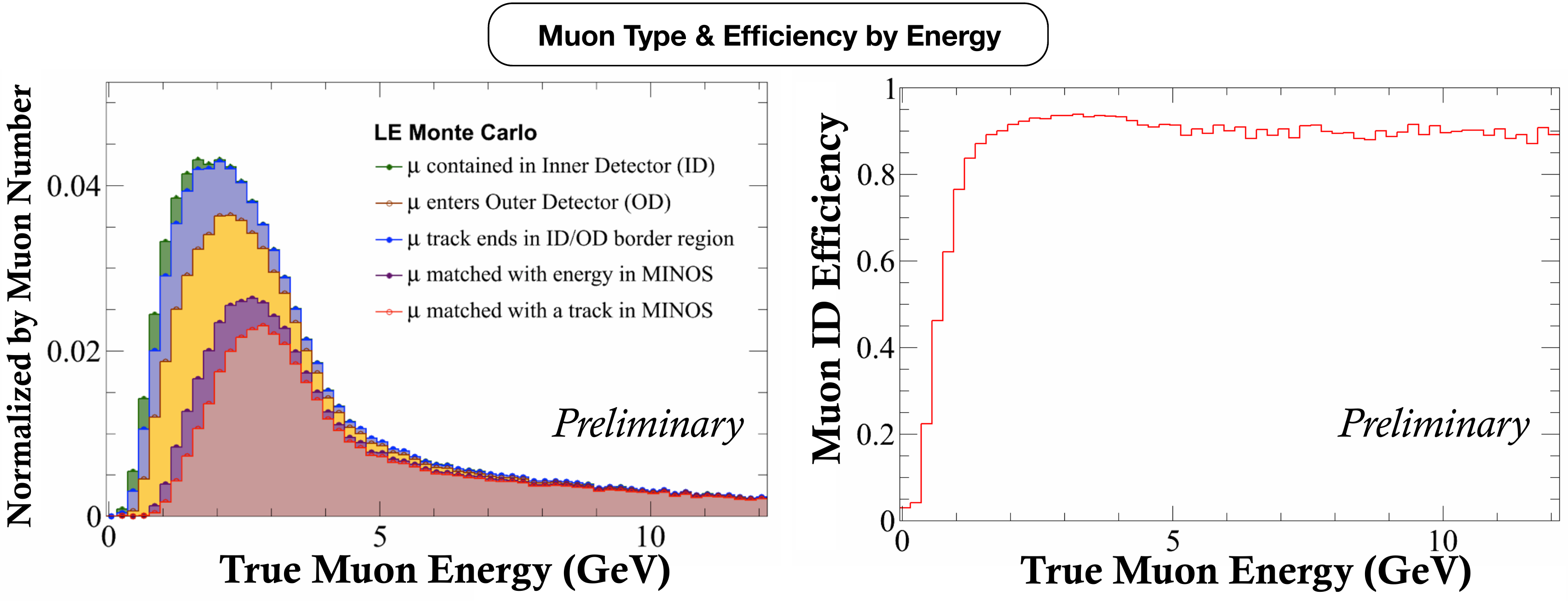}
\end{center}
\caption{\label{fig:MuonEffic}Muon reconstruction topologies and efficiency in $\mnv$. Reconstruction efficiencies for particles matched into MINOS include the effect of reconstruction efficiency in that detector.}
\end{figure}

There are a number of exciting analyses underway in $\mnv$, but early results will focus on charged-current reactions. As such, muon identification is important. Figure \ref{fig:MuonEffic} illustrates muon classification and reconstruction efficiency in neutrino Monte Carlo (MC). $\mnv$ uses GENIE for event generation \cite{genie}. Current reconstruction efforts are focused on muon tracks matched into the MINOS near detector \cite{minos}, and future efforts will begin to emphasize particles stopping in $\mnv$.

\begin{figure}
\begin{center}
\includegraphics[width=20pc]{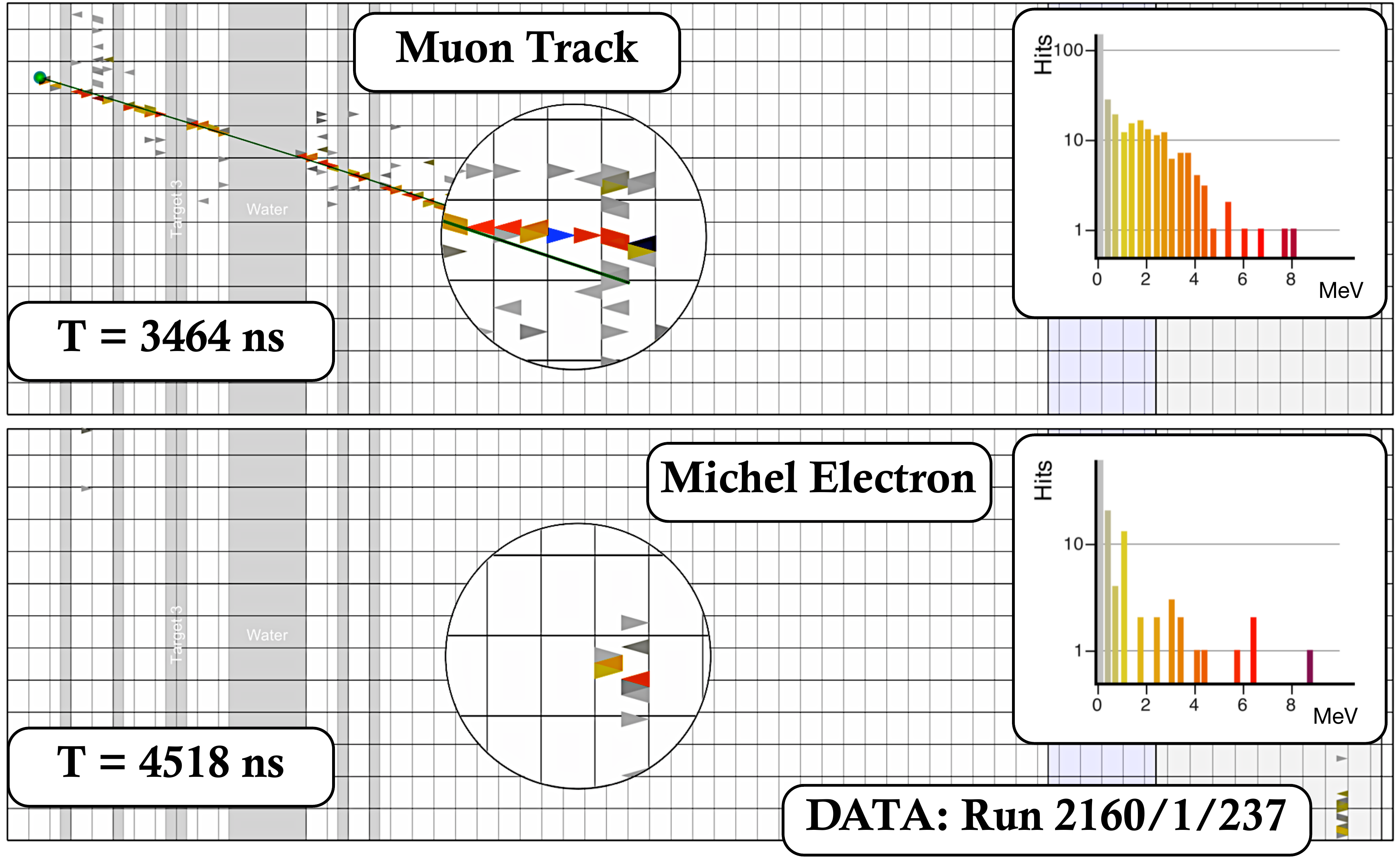}
\end{center}
\caption{\label{fig:MichelDisplay}A Michel electron candidate in $\mnv$ data.}
\end{figure}

\begin{figure}
\begin{center}
\includegraphics[width=28pc]{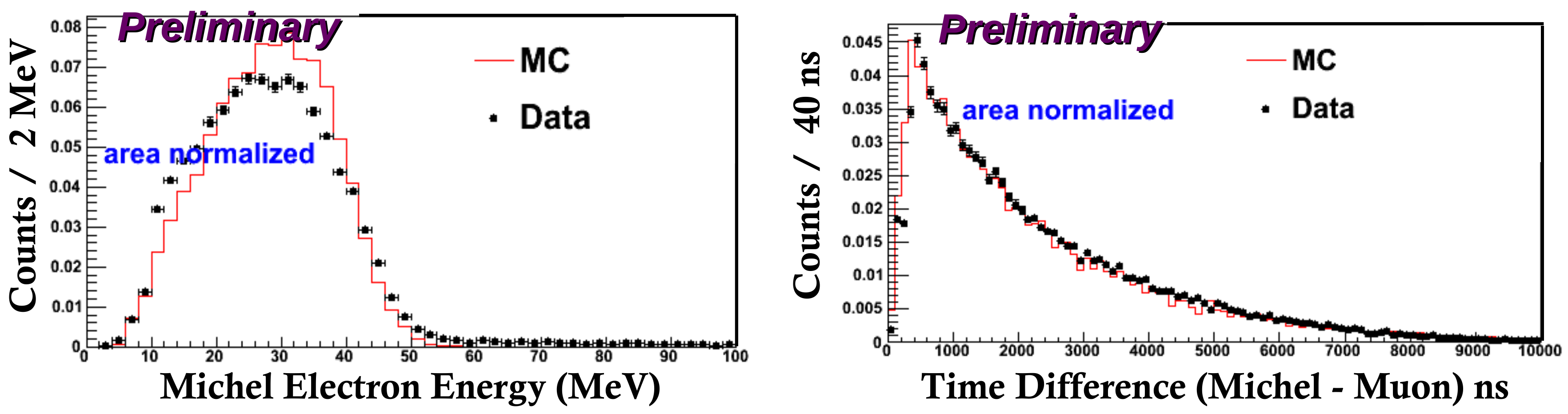}
\end{center}
\caption{\label{fig:MichelFits}A Michel electron energies and lifetimes in $\mnv$.}
\end{figure}

The first step in a stopped muon analysis is Michel electron identification. Figure \ref{fig:MichelDisplay} provides an illustration of an event candidate from data and Figure \ref{fig:MichelFits} is a data to MC comparison of Michel energies and lifetimes in the $\mnv$ tracker. Table \ref{tab:michellife} gives the results of fits to the muon lifetime for the plot on the right in Figure \ref{fig:MichelFits}.

\begin{table}
\caption{\label{tab:michellife}Michel electron lifetime fits. MC is background-free $\mu^-$, while data contains a small $\mu^+$ contamination. The nominal $\mu^-$ lifetime in carbon is 2026 ns.}
\begin{center}
\begin{tabular}{lc}
\br
 & Muon Lifetime in Plastic (ns) \\
\mr
MC   & $2100 \pm 10$ \\
Data & $2120 \pm 20$ \\
\br
\end{tabular}
\end{center}
\end{table}

\section{Conclusion}\label{sec:conclusion}

$\mnv$ is actively collecting data and building software infrastructure for physics analysis. Operations are stable and we are progressing in particle identification and event reconstruction.

\section*{Acknowledgments}
This work was supported by DOE Grant No. DE-FG02-91ER40685.

\end{document}